\documentclass[runningheads,a4paper]{llncs}
\usepackage[utf8]{inputenc}
\usepackage{amsmath}
\usepackage{amsfonts}
\usepackage{algorithm}
\usepackage{algpseudocode}
\usepackage{graphicx}
\usepackage{float}
\usepackage[caption=false]{subfig}
\usepackage{subfig}
\usepackage{CJKutf8}

\algnewcommand\algorithmicinput{\textbf{INPUT:}}
\algnewcommand\Input{\item[\algorithmicinput]}
\algnewcommand\algorithmicoutput{\textbf{OUTPUT:}}
\algnewcommand\Output{\item[\algorithmicoutput]}

\title{A General Method for Event Detection on Social Media}
\author{Yihong Zhang \and Masumi Shirakawa \and Takahiro Hara}
\institute{ Multimedia Data Engineering Lab, Graduate School of Information Science and Technology, Osaka University, Osaka, Japan \\
\email{yhzhang7@gmail.com, shirakawa@hapicom.jp, hara@ist.osaka-u.ac.jp}}
\authorrunning{Zhang et al.}

\begin{document}

\maketitle

\begin{abstract}
Event detection on social media has attracted a number of researches, given the recent availability of large volumes of social media discussions. Previous works on social media event detection either assume a specific type of event, or assume certain behavior of observed variables. In this paper, we propose a general method for event detection on social media that makes few assumptions. The main assumption we make is that when an event occurs, affected semantic aspects will behave differently from its usual behavior. We generalize the representation of time units based on word embeddings of social media text, and propose an algorithm to detect events in time series in a general sense. In the experimental evaluation, we use a novel setting to test if our method and baseline methods can exhaustively catch all real-world news in the test period. The evaluation results show that when the event is quite unusual with regard to the base social media discussion, it can be captured more effectively with our method. Our method can be easily implemented and can be treated as a starting point for more specific applications.
\end{abstract}

\section{Introduction}
Event detection on social media in recent years has attracted a large number of researches. Given large volumes of social media data and the rich information contained in them, event detection on social media is both beneficial and challenging. With social media text as the base data, important previous works have proposed methods for detecting earthquakes \cite{earthquake-twitter-social-sensor}, emerging topics for organizations \cite{chen2013emerging}, and influenza trends \cite{gao2018mapping}. In these works and many others, however, it is required to have some prior knowledge or assumptions of the potential event. These assumptions include some known keywords or entity names that are associated with the event \cite{earthquake-twitter-social-sensor,popescu2010detecting,twitter-topic-detection,chen2013emerging,unankard2014emerging,Olteanu:2014aa}, and some manually created labels for events as the supervised training dataset \cite{Li:2012ab,gao2018mapping}. Furthermore, the definition of event also differs in these works. Some consider an event as a temporal-spatial concentration of similar texts \cite{unankard2014emerging,gao2018mapping,zhou2014event,Li:2012ab,dong2015multiscale}, while others consider it as an unusual burstiness in term frequency \cite{chen2013emerging,weng2011event,rossi2018early}.

In this paper, in contrast, we attempt to provide a general solution to event detection in social media with minimum prior assumption of the event. First of all, we follow a general definition of event that is not restricted to social media data. This definition was proposed by Jaegwon Kim, who considered that an event consists of three parts, a finite set of objects $x$, a property $P$, and a time interval $t$ \cite{kim1976events}. To better illustrate, let us consider a scenario of an amusement park. Normally, customers wander around the park, visiting different attractions in almost a random manner. When a show starts to perform in the central stage, those who are interested in the show will be moving towards the stage. In this scenario, the object $x$ are the customers who interested in the show, the property $P$ is the direction of the stage, and the time interval $t$ is the duration of the show. Note that just as not all customers in the park are interested in the show, $x \in X$ in an event is a subset of all possible objects.

Putting it on the social media case, when an event creates an impact on people's lives, it is likely that it will be reflected on online discussions. Certain semantic aspects of posted text, which can be considered as the object set $x$, would suddenly have unusual trends together, whose deviation can be considered as the property $P$, for the duration of the event $t$. This is realistic, if we recall that when a critical political event happened, some aspects of social media discussion, such as the terms and sentiments involved in the event, would have a higher-than-usual visibility. The problem then is how to capture $x$, $P$ and $t$ in social media text through a computational method.

The principle of our design is to make as few assumptions about the event as possible. Here are two assumptions we make in our method. First, there is a finite set of components in the system, and a subset of components will be affected by the event. Second, for the duration of the event, affected components behave differently from their usual, normal behavior. We consider these are minimum assumptions that are within restrictions in Kim's definition of an event. Given these assumptions, our method takes two steps to achieve event detection. First, we convert unstructured social media text data into distributed representation, also called \emph{word embeddings}, where each dimension represents a semantic aspect, and is considered as a component in the system. This can be done with existing distributed representation learning techniques such as \emph{word2vec} \cite{mikolov2013distributed}. Note that in this paper we consider only social media text. However, the images in social media can be studied in a similar way as they be turned into multi-dimension vector representations using models such as Inception \cite{szegedy2016rethinking}. Second, we design and use a multi-dimension anomaly detection algorithm to capture the unusual behavior, with a customizable normality test. The algorithm detects abnormal intervals in single time series and combines them to form affected components of an event by finding the intersections.

\begin{CJK}{UTF8}{min}
Our method is general in two ways. First, our method generalizes social media text into semantic aspects. With this generalization, we now look at the collective behavior of social media posts instead of tracking individual term frequency. This is useful in many scenarios. For example, during New Year holiday in Japan, many aspects of real-world phenomenon become visible, including New Year's meal (年越し), a specific TV program (紅白), New Year's greeting (挨拶), and the general happy mood. Individually, these terms may not have a significant frequency change, but collectively, they make the New Year event unusual. Second, our method generalizes event detection as anomaly detection in time series. In contrast to previous works, we deal with durative events instead of punctual events. With a customizable normality test function, we can detect events with arbitrary lengths. Such generality allows our method to be applicable to a wider range of tasks than previous works. Since our method is straightforward to implement, future extension can be easily made for the need of specific tasks.
\end{CJK}

We organize the remainder of this paper as the following. In Section 2, we will discuss related works on event detection in social media. In Section 3 and 4, we will present our method to generalize social media text to temporal word embeddings, and to detect unusual behavior in them. In Section 5, we will present experimental evaluation, with a novel evaluation task of recommending newsworthy words. Finally Section 6 will conclude this paper.

\section{Related Work}
Previous surveys on social media event detection works have commonly divided works according to whether detected events are specific or non-specific \cite{atefeh2015survey,saeed2019s}. Here we would like to provide a new aspect of events in existing works, that is whether events are considered as one-time events or events lasting for a period of multiple time units. In other words punctual and durative events. Essentially, punctual events are supposed to be the point of drastic change in the observed variables \cite{guralnik1999event}. While this limits the phenomenon they can represent, events with this definition are indeed easier to capture, and many works followed this approach. For example, the Twitter-based earthquake detection system proposed by Sasaki et al. \cite{twitter-realtime-event-detection} raises an alarm at the moment when number of tweets classified as earthquake reports reaches a certain threshold. Similarly, the event detection system proposed by Zhang et al. raises an alarm at the moment when the number of incident reports within a geographical region reaches a threshold \cite{zhang2018snaf}. Weng and Lee proposed an event detection method based on wavelet transformation and word clustering \cite{weng2011event}. An event flag is set for a time slot if frequency correlation of co-occurring words is larger than a threshold. The crime and disaster event detection system proposed by Li et al. aims to extract the time an event happened, by location estimation and geographical clustering \cite{Li:2012ab}. The location-based event detection method by Unankard et al. also uses a threshold to decide if an event has happened, by comparing the frequency in the current and previous time unit \cite{unankard2014emerging}. The disaster monitor system by Rossi et al. decides if an event happened by determining if word frequency in the current time slot is an outlier \cite{rossi2018early}.

While not uncommon in time series pattern mining  \cite{batal2012mining}, comparing to punctual events, social media event detection methods that follow a durative event definition are rather scarce. Relevant works include the emerging topic detection method proposed by Chen et al., which identifies two time points, the moment the topic starts and the moment the topic becomes hot \cite{chen2013emerging}. The purpose of the method is to identify emerging topic before the topic becomes hot, and detected events thus last for periods of varied lengths. One requirement of the method, however, is that the tweets collected should be related to certain organizations, which makes the method less applicable. The multiscale event detection method proposed by Dong et al. \cite{dong2015multiscale} aims at discovering events with spatio-temporally concentrated tweets. Without a preset time length for the event, the method clusters tweets that have similar spatio-temporal context, and thus indirectly detects events that last for a period. However, the requirement of spatial information also limits the applicability of the method. In this paper, on the other hand, we aim at providing a general method for detecting durative events with less restrictions.

\section{Generalized Representation of Temporal Social Media Text}
We first deal with problem of representing temporal social media text in a general way. A simple way to represent social media text is through bag-of-words (BOW). BOW representation essentially considers that words in text are independent tokens, and each document is a collection of them. There are two problems with BOW representation. First, in a large text collection, the vocabulary is also large, usually includes thousands of words, and tracking temporal activity of each word is computationally expensive. Second, considering words as independent tokens ignores semantic information about words, which may be important for event detection. For example, Covid-19 and Corona are both names of the virus in current pandemic, and should be considered together in one event, but BOW representation would consider them separately.

To mitigate these problems, we propose to use \emph{word embeddings} to represent temporal social media texts. First proposed by Mikolov et al., word embeddings are distributed representation of words learned from text contexts \cite{mikolov2013distributed}. The learning technique extracts the surrounding words of a certain word and encode them in a neural network encoder, so that a vector, called an embedding, can be associated with the word, and each element in the vector represents a certain semantic aspect of the word. While the meaning of the semantic aspect of the embedding is difficult to be understood by human reader, it has been shown that words with similar embeddings would have a similar semantic meaning. For example, \emph{apple} would have a more similar embedding to \emph{orange} than to \emph{bird}.

Using word embeddings thus mitigates the problems of BOW representation. First, it reduces dimensionality. Typical word embeddings would have between 50 and 300 dimensions. Second, it allows consideration of semantics, so that words of similar meanings can be considered together. By considering semantics, we actually generalize text into a more abstract level. For example, when detecting the pandemic event, we no longer deal with individual words such as Covid-19 and Corona, but the virus or disease these words refer to. Given it is effectiveness, previous works have already use word embedding to represent not only text documents, but also users and spatial units such as locations \cite{wang2018reasearch,shoji2018location2vec}. In this work, we utilize word embeddings to generate vector representations of time units.

To generate vector representation for a time unit, we take the following steps. 
\begin{enumerate}
    \item assigning collected text messages to time units.
    \item tokenizing text messages so that words are also assigned to time units
    \item obtaining word embeddings for assigned words
    \item the vector representation for a time unit is taken as the average value of all embeddings of the words assigned to the time unit
\end{enumerate}

We can use existing natural language processing libraries to segment and turn tweets into words. To obtain word embeddings, we can use existing implementations of \emph{word2vec} and a general purpose training corpus such as Wikipedia\footnote{An example online resource that provides an implementation under this setting: https://github.com/philipperemy/japanese-words-to-vectors}. Word embedding learned under such setting would represent words with their general meaning in daily usages. The final result of this process is a vector representing the totality of social media discussions for each time unit.

\section{Generalized Multi-dimension Event Detection in Time Series}
At this point we have a vector for each time unit representing social media discussions. The next task is to detect events from such representations. In a way this representation can be seen as multivariate time series data, with each dimension as one observed variable. While there are previous works that have proposed event detection for time series data, most of them are dealing with punctual event \cite{guralnik1999event,cheng2009detection}, or require the events to be repeating and predictable \cite{batal2012mining}. In this work, we accept the hypothesis that an event is something that cannot be predicted, thus the behavior of affected components cannot be pre-defined \cite{taylor2009black}. We aim to make minimum assumptions about the event, and the main assumption we make is that when affected by an event, the component will behave differently from its usual behavior.

Our method detects multi-dimension event from multivariate time series in two steps. First it detects unusual intervals of observations in a single dimension (Algorithm \ref{alg:single}). Then given a list of abnormal intervals in each dimension, it finds basically the intersections of abnormal intervals, and outputs them as multi-dimension events (Algorithm \ref{alg:multi}). 

\begin{algorithm}[ht]
\caption{Find largest intervals with significant alternation to normality}
\label{alg:single}
\begin{algorithmic}[1]
\Input $TS$, $k_{min}$, $k_{max}$, $f_n$, $\delta$
\Output a list of intervals $Is$
\State $Is \gets \{\}$
\State $i \gets 1$
\While {$i < (|TS| - k_{min})$}
\State $largest\_interval \gets \{\}$
\For {$j$ in $(i+k_{min})$ to $min(|TS|, i+k_{max})$}
\If{$f_n(TS \setminus TS(i, j)) - f_n(TS) > \delta$}
\State $largest\_interval \gets (i, j)$
\EndIf
\EndFor
\If{$largest\_interval$ is empty}
\State $i \gets i+1$
\Else
\State $Is \gets Is \cup largest\_interval$
\State $i \gets (b \text{ in } largest\_interval) + 1$
\EndIf
\EndWhile
\end{algorithmic}
\end{algorithm}

Shown in Algorithm \ref{alg:single}, we design an algorithm to find the largest interval with significant alternation to normality. It takes a univariate time series as input, as well as two parameters $k_{min}$ and $k_{max}$, which are the minimum and maximum number of time units for the detected intervals. It also requires a customizable function $f_n$ for the normality test, and a corresponding threshold $\delta$. The algorithm starts from the beginning of the time series (line 2, 3). At each time point $i$, it tests all intervals that ends between $i + k_{min}$ and $i + k_{max}$ (line 5). With each interval, it performs normality test with the specified function $f_n$, and if the normality difference between the time series with and without the interval is larger than $\delta$, then the interval is considered abnormal (line 6). The largest interval considered as abnormal will be taken as the abnormal interval starts at time $i$ (line 7). If an abnormal interval is found, the algorithm will move to the end of the interval (line 13, 14), and continue until it reaches the end of the time series. Finally the algorithm returns all abnormal intervals found as $Is$.

\begin{algorithm}[t]
\caption{Find multi-dimension events}
\label{alg:multi}
\begin{algorithmic}[1]
\Input $Is, k_{min}, c_{min}$
\Output $E$
\State $E \gets \{\}$
\State $E_{half} \gets \{\}$
\For{$i$ in 1 to $n - k_{min}$}
\State $D_{cur} \gets \{d_j| i \in Is_j\}$
\State $D_{old} \gets \{\}$
\For{each $e_{half} \in E_{half}$}
\State $D_{continuing} \gets d(e_{half}) \cap D_{cur}$
\If{$D_{continuing} = \{\}$}
\State next
\EndIf
\State remove $e_{half}$ from $E_{half}$
\If{$|D_{continuing}| > c_{min} $}
\State $e_{continuing} \gets (start(e_{half}), i, D_{continuing})$
\State $E_{half} \gets E_{half} \cup e_{continuing}$
\State $D_{old} \gets D_{old} \cup D_{continuing}$
\Else
\State $D_{continuing} \gets \{\}$
\EndIf
\State $e_{finished} \gets (start(e_{half}), i, d(e_{half}) \setminus D_{continuing})$
\If{$l(e_{finished}) > k_{min} \And |d(e_{finished})| > c_{min}$}
\State $E \gets E \cup e_{finished}$
\State $D_{old} \gets D_{old} \cup (d(e_{finished}) \cap D_{cur})$
\EndIf
\EndFor
\State $D_{new} \gets D_{cur} \setminus D_{old}$
\If{$|D_{new}| > c_{min}$}
\State $E_{half} \gets E_{half} \cup (i, i, D_{new})$
\EndIf
\EndFor
\end{algorithmic}
\end{algorithm}

It is worth noting that Algorithm \ref{alg:single} does not necessarily find intervals that deviate most from normality. For example, given a highly abnormal interval $I$, a few time units surrounding $I$ may be normal by themselves, but when considered together with $I$, this larger interval may still be abnormal above the threshold. And our algorithm will pick the larger interval instead of the more deviating interval. Since our goal is to detect multi-dimension events, and the intervals are to be taken as the input of next step, it is rather desirable to have the largest possible abnormal intervals, instead of smaller, more deviating intervals.

The normality test function $f_n$ can be defined by the user, as long as it outputs a score for data normality or randomness. There are many existing normality test functions available to use, including Box test and Shapiro Wilk test \cite{bartels1982rank}. For the completion of the method, we use the rank version of von Neumann’s ratio test \cite{bartels1982rank} in our experimental analysis \footnote{An implementation of this test is available as an R package: https://cran.r-project.org/web/packages/randtests/randtests.pdf}. After some trying a few test functions, we found that this randomness test tests to capture unusual intervals in data more consistently.

After processing the data with Algorithm \ref{alg:single}, we now have a list of abnormal intervals $Is$ for each of the word embedding dimension. The goal of next algorithm, shown as Algorithm \ref{alg:multi}, is to find the intersection of these intervals. It is an incremental algorithm that needs to go through the dataset only once. It takes the set of $Is$ as inputs, as well as two parameters, $k_{min}$ as the minimum length of an event period, and $c_{min}$ as the minimum number of affected dimensions in an event.

At each time point $i$, the first thing to do is find the dimensions that behave unusually at $i$, based on the intervals detected (line 4). From there, these dimensions are either considered as a part of a continuing event, or put to form a new event. We always keep a list of events that are halfway through $E_half$, and at each time point, we check through all halfway events for continuity (line 2, 6). If affected dimensions at time $i$ match halfway events, they are assigned to these events, and if enough dimensions are assigned ($> c_{min}$), the halfway event is considered as continuing (line 7 to 18). If a halfway event could not be matched with enough affected dimensions, the event is considered as finished (line 19 to 23). Those dimensions not matched with any halfway event are grouped to form a new halfway event, if there are enough of them (line 25 to 28). The final output is a list of events $E$, where each $e \in E$ has $e=\{\mathbf{x}, t\}$, with $\mathbf{x}$ as affected dimensions, and $t$ as the event period.

\section{Experimental Evaluation}
We use real-world social media data to verify the effectiveness of our event detection method. We are unable to establish a way to directly evaluate the detected events, which consist of duration and affected dimensions in word embeddings, and are not human-readable. Therefore, we attempt to evaluate them indirectly. We extend our method to perform a task called recommending newsworthy words, which has been the evaluation task in other event detection works \cite{unankard2014emerging,dong2015multiscale}. We will present the details of this task and the results in this section. It is worth noting here, though, that our event detection can potentially do more than recommending newsworthy words.

\subsection{Evaluation Task}
Our evaluation task is as follows. Given a set of time units $T=\{t_1,...,t_c\}$, for each time unit, we apply the event detection method on a social media discussion dataset, and generate a ranked list of event words $P$ from detected events. Also for each time unit, we generate from news sources a ranked list of news words $G$. The evaluation is done by comparing $P$ and $G$. If $|G \cap P|$ is large, then the event detection method is considered as capable of capturing newsworthy words, which also shows that the news has an impact on the social media discussion.

Traditional evaluation of event detection is centered on detected events \cite{unankard2014emerging}. It verifies whether detected events is corresponding to real-world events, and does not do anything when a real-world event has not been detected (false negative). We on the other hand, attempt an exhaustive evaluation that concerns all real-world events happened. Specifically, we consider all news headlines from news source for each time unit, and evaluate to what degree corresponding information can be detected by the event detection method.

\subsection{Social Media Discussion Dataset}
Since it is not feasible to monitor all messages in a social media platform such as Twitter, we select a subset of all messages on Twitter as our social media discussion dataset. First we obtain a list of Japanese politician Twitter accounts\footnote{Since politician are public, such a list can be found in many online sources, for example: https://meyou.jp/group/category/politician/}. Then we monitor all tweets mentioning these accounts using Twitter Stream API\footnote{https://developer.twitter.com/en/docs/tutorials/consuming-streaming-data|}. For period of six months between January and July, 2020, we collected about 6.9 million tweets, after removing retweets. We take this as the discussion dataset. We understand this dataset does not represent the overall discussion happening on Twitter, but rather has a focused theme that is Japanese politics. But such discussions and the community producing them may still be affected by general news, and it will be interesting to see what unusual events can be captured from these discussions and how they correspond to news sources. It is expected that if we can detect the events in this discussion dataset, we can also detect events in discussion of different themes in the same way.

We use the natural language processing package kuromoji\footnote{https://github.com/atilika/kuromoji} to process the Japanese text in social media discussion dataset. The package can effectively perform segmentation and part-of-speech (POS) tagging for Japanese text. After POS tagging, we select only nouns to represent the information in the text. We also filter out some less frequent words, and consider only 8,267 words that have appeared at least 500 times in the dataset.

\subsection{Ground Truth Generation}
We generate ground truth news words as follows. First we collect messages posted by a number of Japanese news Twitter accounts\footnote{A list of popular Japanese news Twitter accounts can be found on the same source: https://meyou.jp/ranking/follower\_media}. Among 916 news account considered, some are general news accounts reporting local and international news, some are specific news accounts reporting news for example in sports or entertainment. Messages sent from these accounts are usually news headlines. To make our target clearer, we select from collected messages three specific topics, namely, \emph{politics}, \emph{international}, and \emph{Corona}. The selection is done by filtering collected messages with these three topic words as hashtags. During a one-month period between June and July, 2020, we collected 814 political news headlines, 503 international news headlines, and 602 Corona news headlines. These news headlines are assigned to time units of one hour length.

We turn these news headlines into nouns by the same kuromoji software described in the previous section, and count the frequencies. These words are then ranked using $tfidf$, which is calculated as:
\[
tfidf(w) = tf(w) \cdot \log \frac{|D|}{|d \in D : w \in d|}
\]
where $tf(w)$ is the frequency of word $w$, and $D$ is a collection of documents, which in our case is messages assigned to $|D|$ time units. Finally, for each time unit, we pick top-20 words ranked by $tfidf$ as the ground truth news words.

\subsection{Recommending Newsworthy Words from Detected Events}
Since our method does not generate ranked words directly, we need a method to convert the output of our method into words. The output of our method is a list of events $E=\{e_1,...,e_m\}$, where for each event we have a set of affected dimensions $\mathbf{x}$ and duration $t$.

To convert this result back to words, we first calculate the deviation of a affected dimension in the event duration as the difference between mean value of the dimension in the event duration, and the mean value outside the duration:
\[
dev_e(x) = mean\_freq(x, t) - mean\_freq(x, \neg t)
\]
which can be considered as a part of event property $P$. Then for each word $w$ with embedding $embedding_w$, an event score is calculated as the product of the embedding value and the deviation in the affected dimensions:
\[
event\_score_e(w) = \sum_{x \in \mathbf{x}} embedding_w(x) \times dev_e(x)
\]
In this way, words with the same deviation tendency as the affected dimensions will have a higher score. Finally, to calculate a word score in a time unit, we have
\[
time\_score(w) = \sum_{e=1}^m event\_score_e(w)
\]
which gives higher scores to words with higher event scores in multiple events. The time score is thus used to rank the words in each time unit.

\subsection{Baseline Methods}
We compare our method with two baseline methods in this evaluation task. The first is a $tfidf$-based method commonly used in previous works. In the same way we generate ground truth, we apply the method to the social media discussion dataset and obtain a $tfidf$ score for each word in each time unit. Essentially, with this method, we make a comparison of $tfidf$-ranked words between base source, which are social media discussion tweets, and the reference source, which are news tweets.

The second baseline method is based on the Shannon's Wavelet Entropy (SWE). This method is proposed in a Twitter event detection work by Weng and Lee \cite{weng2011event}, and can be adopted for news word recommendation. From the $tfidf$ time series of each word in the social media discussion dataset, the method first performs a wavelet transformation to learn a wavelet function $\psi$ and a coefficient $C$. The coefficient $C$ can be interpreted as the local residual errors. Then an energy value $E$ is calculated as
\[
E = \sum_k |C(k)|^2
\]
where $k$ indicates $k$-th coefficient. Then the Shannon's Wavelet Entropy is calculated as
\[
SWE = - \sum_j \rho_j \cdot \log \rho_j
\]
where $\rho_j = E_j / E_{total}$, $j$ indicates the $j$-th time unit in the time slide. SWE measures how unpredictable of the time series in a time slide $t$, and it will be a higher value when residual errors are more even in the time slide. Once the SWE is obtained, a score can be assigned to a word for ranking.
\[
s(w) =
\begin{cases}
\frac{SWE_t - SWE_{t-1}}{SWE_{t-1}}, &\text{if } SWE_t > SWE_{t-1}\\
0, & \text{otherwise}
\end{cases}
\]
which means if SWE of a word is increasing, it will get a higher score. In our experiments, we use the R package \emph{wavethresh}\footnote{https://cran.r-project.org/web/packages/wavethresh/wavethresh.pdf} to perform the wavelet transformation and obtain coefficient $C$.

\subsection{Evaluation Results}
\begin{figure}[ht]
\centering
\vspace{-30pt}
\subfloat[political]{\includegraphics[width=0.47\textwidth]{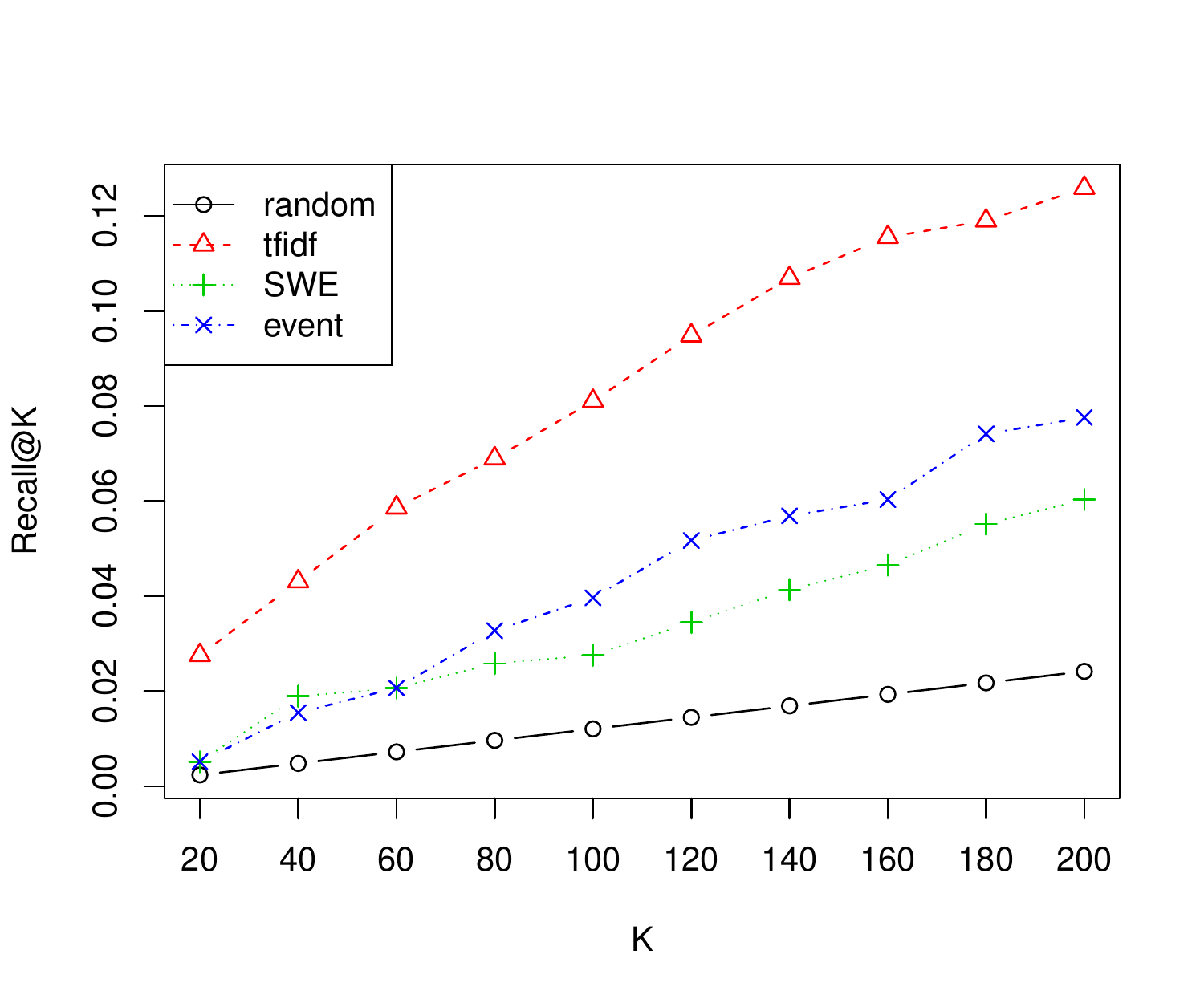}} 
\subfloat[international]{\includegraphics[width=0.47\textwidth]{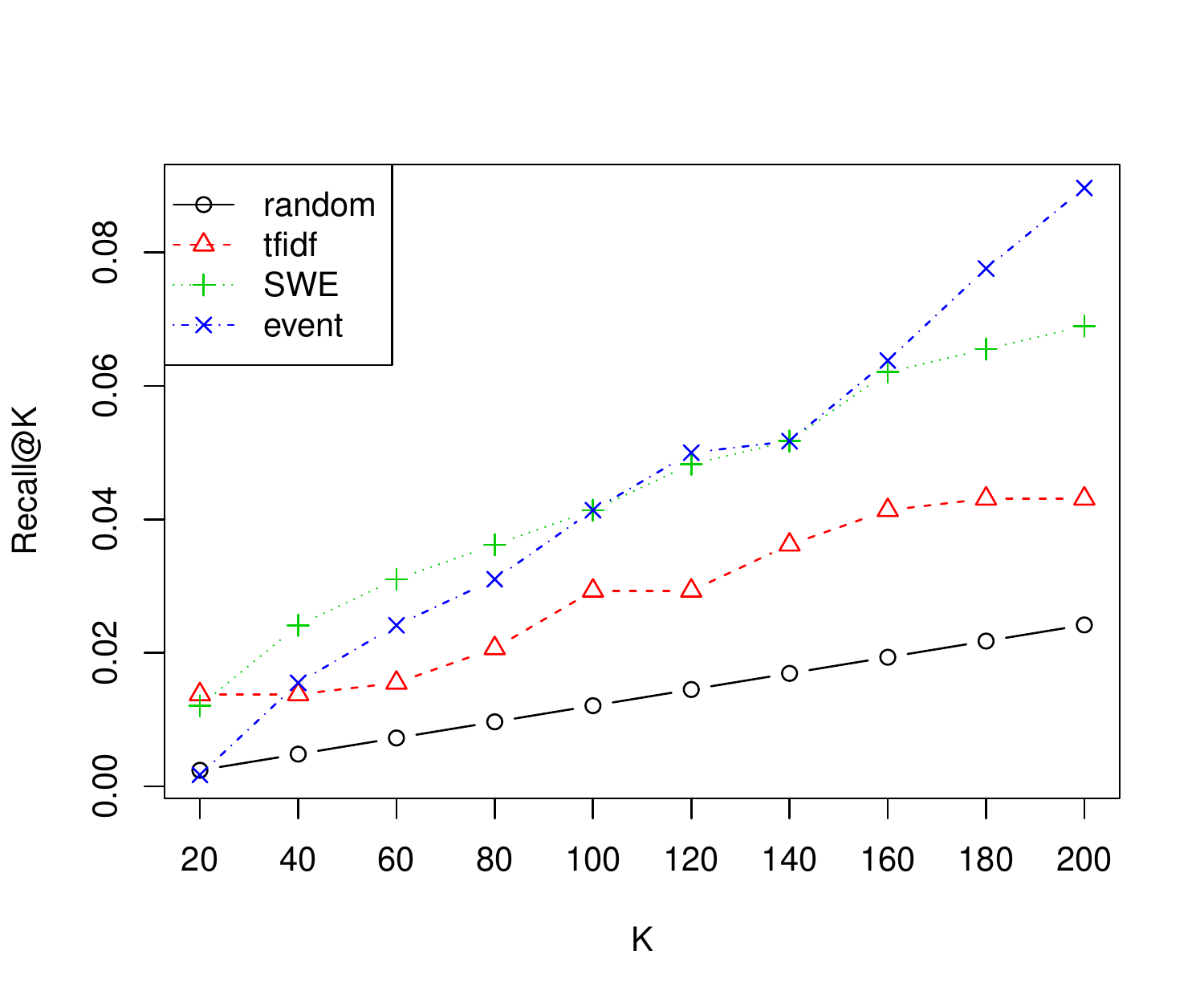}}\\
\vspace{-20pt}
\subfloat[Corona]{\includegraphics[width=0.47\textwidth]{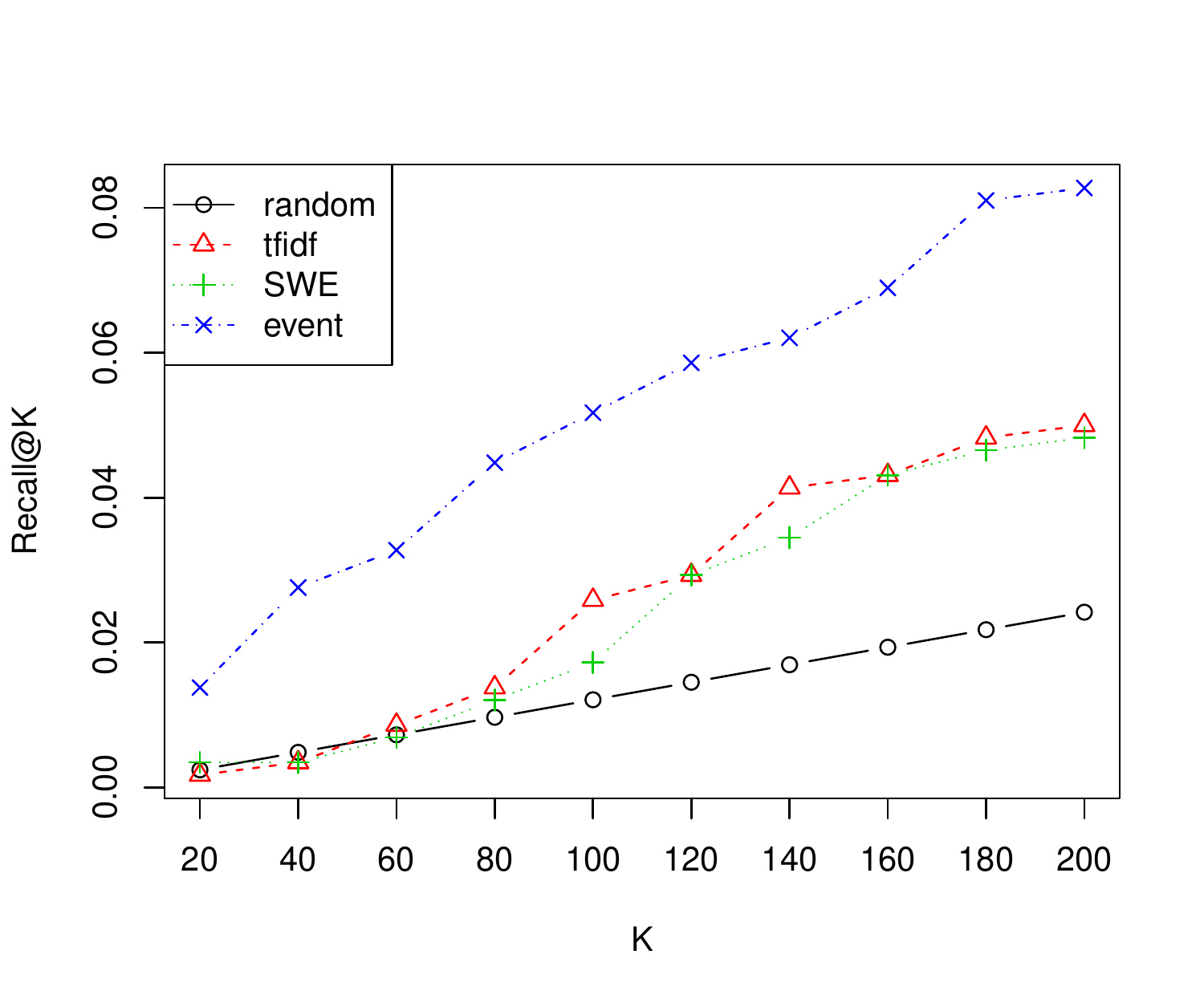}} 
\caption{Recall@K results for three news categories}
\label{fig:acc}
\end{figure}

Evaluation results measured as Recall@K are shown in Fig. \ref{fig:acc}, where K is the number of recommended words. We compare our method (event) with theoretic random, \emph{tfidf}, and SWE methods. A number of Ks are taken between 20 and 200. The theoretical random Recall@K is  calculated as $K/|W|$, where $W$ is the set of candidate words. The higher the result means the more words in ground truth are recommended by the method.

At the first glance, we can see that generally, \emph{tfidf} performs better for the political news, while event method performs better for the Corona news. SWE method performs better for the international news, although only slightly better than the event method. All three methods achieve better results than the theoretical random method.

We now attempt to explain the results. First thing to note is that recommended words from a method is the same for all three news categories. Since words from news categories are quite different, with limited space, a method better at recommending words for one news category will be worse for other categories. And we can see the results are showing different strength and weakness from different methods. The reason comes from different interpretation of what is news by different methods. For the \emph{tfidf} method, news is considered as unusually rises of word usages, and thus words closer to the theme of the social media discussion will be more likely to be recommended. For the event method, news is considered as something quite different from the usual state of the discussion, and thus words different from the social media discussion will be more likely to be recommended. And indeed we understand that, since the social media discussion is generally related to politics, political news is more similar to the discussion, while Corona news is more different from the discussion. That is why we see \emph{tfidf} performing better for political news, and event method performing better for Corona news.

\section{Conclusion}
In this paper we propose a general method for event detection on social media. Two main steps of our method are generalizing social media text into word embeddings, and detecting multi-dimension event from time series. The detected events represent something unusual and affecting semantic aspects of social media discussions, over a finite period. Comparing to previous works on social media event detection, our method makes very few assumptions. We only assume that the event will be affecting a finite number of dimensions and, when affected, these dimensions behave differently from their usual, normal behavior. We evaluate detected events from social media discussions against three news categories, exhaustively collected over a testing period, and find that when the news is quite different from the base social media discussion, it can be better captured based on the detected events.

Despite some positive results from indirect evaluation, we consider that our method has some drawbacks. For example, our method demands test of normality, and requires a large portion of base data, which may not be always available. Furthermore, if it is a long period event, event-related semantics would become the norm and thus there would be problem detecting the event with our method. Nevertheless, our method has its merits. It can be easily implemented and applied to different, more specific datasets. One can, for example, pre-select a discussion dataset about finance or entertainment, and apply our method to detect events of certain type. The detected events can furthermore be used in various analysis, for example, for detecting associations between product sales and social media activities. Currently, our method detects events retrospectively. A future extension to our method would be making an incremental algorithm that can detect events in data streams.

\section*{Acknowledgement}
This research is partially supported by JST CREST Grant Number JPMJCR21F2.

\end{document}